# Spherically Symmetric Considerations for a Higher Order Theory of Gravitation

S N Pandey and B.K.Sinha

**Abstract:**

A higher order theory of gravitation is considered which is obtained by modifying Einstein field equations. The Lagrange used to modify this in the form a polynomial in (scalar curvature) R. In this equation we have studied spherical symmetric metric.

## Introduction:

It has been known for long time that there is nothing absolute in this universe. The relative motion as observed in the universe has given much principle that has accounted for by Einstein in the form of equivalence principle to a great extent. Therefore, to understand the true character of this interplay the solution of Einstein's equations are very essential. We attempted to develop a higher-order theorem of gravitation based on the conformal non invariance of gravitational wave equations. This has done because gravitational radiations are an inevitable consequence of Einstein theory of gravitation and they are meaningfully comparable with electromagnetic waves except, at least, in this conformal non-invariance. This motivation led us to modify the Einstein theory (based on Hilbert Lagrangian, R) by considering a Lagrangian density [1, 2]

$$h_g = \sqrt{-g}\left[R - \sum_{n=2}^{N} c_n \frac{(l^2 R)^n}{6l^2}\right] \qquad (1)$$

Here $c_n$ are arbitrary dimensionless coefficients corresponding to n, introduced to nullify the manifestations of gravitation in the form of $\frac{a''}{a}$ in the graviton equation [3, 4]

$$\mu'' + \mu\left[n^2 - \frac{a''}{a}\right] = 0 \qquad (2)$$

Where a is the scale factor of Friedmann universe taken in conformal form

$$ds^2 = -g_{\mu\nu}dx^\mu dx^\nu = a^2(\eta)(d\eta^2 - dx^2 - dy^2 - dz^2), \quad (3)$$

$\eta$ is related to cosmic time t by the relation

$$cdt = a(\eta)d\eta \quad (4)$$

And $a' = \dfrac{da}{d\eta}$. The Lagrangian (1) contains a polynomial in R of a finite number of terms. This should not be disturbing because it is an observational fact that our universe is not asymptotically flat. There is enough matter on our past light cone to cause it to refocus. The total energy of the universe is exactly zero, the positive energy of gravitation and the matter particles being exactly compensated by the negative gravitational potential energy. That is why the universe is expanding. Also the unitarity is not well defined, except in scattering calculations in asymptotically flat spaces. Therefore, we consider the above Lagrangian (2) in the form

$$L = R + \sum_{n=2}^{N} a_n R^n \quad (5)$$

For the field equation of higher order theory of gravitation (5) which we study to fuid the gravitational field surrounding a spherically symmetric mass distribution at rest and also the equation of motion of a particle. Here $a_n$ is the coefficient corresponding to n, and n cannot assume values 0 and 1 because in the cases the Lagrangian (5) will reduce down to Hilbert Lagrangian.

## 2. FIELD EQUATION:

An application of the variational particle to the action

$$A = \int \left( \dfrac{L}{K} + L_s \right) d^4 x$$

Where L is given by (5) and $L_s$ stands for the source Lagrangian density, gives the field equations of the "modified field theory" or a higher-order theory as

$$R_{\mu\nu} - \dfrac{1}{2} g_{\mu\nu} R + \sum_{n=2}^{N} n a_n R^{n-1} \left[ R_{\mu\nu} - \dfrac{1}{2n} g_{\mu\nu} R - \dfrac{n-1}{R}(R_{;\mu,\nu} - g_{\mu\nu}\Box R) - \dfrac{(n-1)(n-2)}{R^2}(R_{,\mu}R_{,\nu} - g_{\mu\nu}R_{,\alpha}R^{,\alpha}) \right] = kT_{\mu\nu} \quad (6)$$

Here

$$T_{\mu\nu} = \sqrt{-g}\,\frac{\delta L_s}{\delta g^{\mu\nu}} \tag{7}$$

Stands for the energy-momentum tensor responsible for the production of the gravitational potential $g_{\mu\nu}$.

It can be easily be seen that

$$T^{\nu}_{\mu;\nu} = 0 \tag{8}$$

Holds for (6), as in general relativity. It is evident that the correction term $\sum_{n=2}^{N} a_n R^n$ in (5) has resulted in appearance of terms containing $\sum_{n=2}^{N} n a_n R^{n-1}$ in equation (6). It should be noted that $1 + n a_n R^{n-1} \neq 0$, or equivalently,

$$1 + 2a_2 R + 3a_3 R^2 + 4a_4 R^3 + \ldots + N a_N R^{N-1} \neq 0 \tag{9}$$

because of the Cauchy problem. This fact is important in studying the completeness of geodesic in higher-order gravity theory.

## 3. Spherically Symmetric Field

We want to study the gravitational field surrounding a spherically symmetric mass distribution at rest. It is obvious that this gravitational field should also have spherical symmetry. We require the field to be "Static". This means that it is both time independent and time symmetric that is, unchanged by time reversal. Therefore, we consider the interval as

$$ds^2 = e^{N(r)} dt^2 - e^{L(r)} dr^2 - r^2 d\theta^2 - r^2 \sin^2\theta d\phi^2 \tag{10}$$

The unknown function N(r) and L(r) are to be determined by using the modified field equation (6). The only non-vanishing components of $R^{\nu}_{M}$ are diagonal components. They are

$$G_0^0 = R_0^0 - \frac{1}{2}R = -e^{-L}\left(\frac{L'}{r} - \frac{1}{r^2}\right) - \frac{1}{r^2}$$

$$G_1^1 = R_1^1 - \frac{1}{2}R = e^{-L}\left(\frac{N'}{r} - \frac{1}{r^2}\right) - \frac{1}{r^2}$$

$$G_2^2 = R_2^2 - \frac{1}{2}R = e^{-L}\left(\frac{N'}{2} - \frac{L'N'}{4} + \frac{N'^2}{4} - \frac{N'-L'}{2r}\right)$$

$$G_3^3 = R_3^3 - \frac{1}{2}R = e^{-L}\left(\frac{N''}{2} - \frac{L'N'}{4} + \frac{N'^2}{4} - \frac{N'-L'}{2r}\right)$$

(11)

Higher order field equation (6) in vacuum, that is, $T_{\mu\nu} = 0$, for the space time (10) fields:

$$R_0^0 - \frac{1}{2}R = R_1^1 - \frac{1}{2}R = R_2^2 - \frac{1}{2}R = R_3^3 - \frac{1}{2}R = f(R) \tag{12}$$

Where

$$f(R) = \frac{\sum_{n=2}^{N} a_n\left(2 - \frac{n}{2}\right)R^n}{1 + \sum_{n=2}^{N} na_n R^{n-1}} \tag{13}$$

To obtain the function N(r) and L(r), we solve $R_0^0 - \frac{1}{2}R = R_1^1 - \frac{1}{2}R$

By substituting the values from (11). This gives

$$N(r) = -L(r) \tag{14}$$

Now we consider equation

$$R_0^0 - \frac{1}{2}R = f(R)$$

Putting values from (11) and integrating we find

$$e^{-L} = 1 + \frac{Const}{r} + \frac{1}{r}\int r^2 f(R) dr \tag{15}$$

The "constant" can be determined from the fact that at large distances $g_{00}$ component of the metric (10) must confirm to Newtonian potential, that is, $1 - 2\phi$, if M is the central mass there

Const= -2MG

Letting to metric (10), in view of (14) and (15) as:

$$ds^2 = \left(1 - \frac{2MG}{r} + \frac{1}{r}\int r^2 f(R)\right)dt^2 - \frac{dr^2}{\left(1 - \frac{2MG}{r} + \frac{1}{r}\int r^2 f(R)\right)} \quad (16)$$
$$- r^2 d\theta^2 - r^2 \sin^2\phi d\phi^2$$

It is to be noted that the mass M is the total mass of the system. The mass-energy contributed by the gravitational field is to be included in M. clearly, in view of principle of equivalence, the gravitational mass of the system which produces the field (16) is, in fact, equal to the initial mass of the system.

The other equations in (12) are satisfied by (14) and (15).

Now we turn our attention towards equation (13). The denominator of (13) in view of (9) is non-vanishing. Here f(R) can be expanded in power of r. thus

$$f(R) = b_0 + b_1 r + b_2 r^2 + b_3 r^3 + \ldots \quad (17)$$

Obviously f(R) is a very small contribution. However

$$\frac{1}{r}\int r^2 f(R) dr = \frac{b_0 r^2}{3} + \frac{b_1 r^3}{4} + \frac{b_2 r^4}{5} + \ldots \quad (18)$$

If we retain the first term only, then we have

$$e^N = e^{-L} = 1 - \frac{2MG}{r} + \frac{b_0 r^2}{3} \quad (19)$$

And the space time metric (10) becomes

$$ds^2 = \left(1 - \frac{2MG}{r} + \frac{b_0 r^2}{3}\right) dt^2 - \frac{dr^2}{\left(1 - \frac{2MG}{r} + \frac{b_0 r^2}{3}\right)} - r^2 d\theta^2 - r^2 \sin^2\phi d\phi^2 \quad (20)$$

The metric (20) does not become asymptotically flat as $r \to \infty$, because of contribution $\frac{b_0 r^2}{3}$ coming from the modification of Einstein theory. This behaves like a contribution that comes from cosmological constant if Einstein theory was considered with cosmological constant, that is, $R_\mu^\nu - \frac{1}{2}\delta_\mu^\nu R = -\Lambda \delta_\mu^\nu$. This contribution is small.

In particular, if $G = 1$ & $b_0 = 0$

This follows the wellknown Schwarzscild metric

$$ds^2 = \left(1 - \frac{2M}{r}\right)dt^2 - \frac{dr^2}{\left(1 - \frac{2M}{r}\right)} - r^2\left(d\theta^2 + \sin^2\phi d\phi^2\right)$$

Where M is the total mass of sphere given by $M = 4\pi \int_0^{R_b} \rho(r)r^2 dr$, where $R_b$ is the Radius of the fluid sphere.

If $\frac{1}{\sqrt{b_0}} \gg r \gg GM$, the metric (20) is nearly flat. The effect of mass term M dominates

For the values of r below this range and the effect of this term $\frac{b_0 r^2}{3}$ dominates for the values of r above this range. However in this sceneries, the Newtonian potential gets modified to $\phi = -\frac{GM}{r} + \frac{b_0 r^2}{6}$. The second term appears due to the correction in Hilbert Lagrangian to modify Einstein theory. The contributions of remaining terms in (17) are considerably negligible by definition.

Again it is worth mentioning that the scalar curvature R in Higher-order theory (6) behaves as a wave. This is not so in Einstein theory. For instance, let us take T=0 (vacuum) and n=2, the trace of (6) gives R - $\frac{1}{6a_2}$R =0, which is a wave equation comparable with mass less scalar field equation $\phi + \frac{R}{6}\phi = 0$. R is non vanishing for all values of $n \geq 2$.

## Conclusion:

This shows that f(R) in equation (13) is of wave nature and its expansion (17) appears logically valid. Further, if f(R) is a constant or zero then (16) will correspond to Schwartzschild solution of Einstein field equations with or without cosmological constant. This shows the effect of modification of Einstein field theory based on purely by gravitational field as evident by Lagrangian (5).

To obtain conform ally invariant gravitational wave equations may succeed to a point but in case of spherically symmetric distribution of matter at rest does not contribute significantly so as to produce some observable effects. One of the possible reasons seems to be the fact that Lagrangian (5) depend only on scalar curvature R and is not associated with other fields like scalar field, meson field and like.

**Acknoledgement-Authors would like to thanks Inter university ceter for astronomy and Astrophysics for giving the facility of reprints through speedpost on request of email.Also thanks to MNNIT and HCST college admin .Finally this paper is devoted to Lord Shiva(BHOLE BABA).**

Author Adress:1-MNNIT,Allahabad(U.P.)-India
            2-depott.of Applied Maths,HCST,Farah,Mathura(U.P.)-India